\documentclass[conference]{IEEEtran}

\IEEEoverridecommandlockouts
\usepackage{tikz}
\usepackage{textcomp}
\usepackage{hyperref}
\usepackage{lipsum}

\newcommand\copyrighttext{%
  \footnotesize \textcopyright 2025 IEEE. Personal use of this material is permitted.  Permission from IEEE must be obtained for all other uses, in any current or future media, including reprinting/republishing this material for advertising or promotional purposes, creating new collective works, for resale or redistribution to servers or lists, or reuse of any copyrighted component of this work in other works.}
\newcommand\copyrightnotice{%
\begin{tikzpicture}[remember picture,overlay]
\node[anchor=south,yshift=10pt] at (current page.south) {\fbox{\parbox{\dimexpr\textwidth-\fboxsep-\fboxrule\relax}{\copyrighttext}}};
\end{tikzpicture}%
}

\usepackage{cite}
\usepackage{amsmath,amssymb,amsfonts}
\usepackage{algorithmic}
\usepackage{graphicx}
\usepackage{textcomp}
\usepackage{xcolor}
\usepackage{lipsum}
\usepackage{float}
\usepackage{diagbox}
\usepackage{balance}
\usepackage{array}
\usepackage{footnote}
\usepackage{booktabs}
\usepackage{adjustbox}
\usepackage{multirow}
\usepackage[per-mode=symbol]{siunitx}
\def\BibTeX{{\rm B\kern-.05em{\sc i\kern-.025em b}\kern-.08em
    T\kern-.1667em\lower.7ex\hbox{E}\kern-.125emX}}
\newcommand{\lt}{\ensuremath <}

\makeatletter
\renewcommand{\@makefntext}[1]{%
  \parindent 0pt % Remove indentation
  \noindent
  \@makefnmark#1%
}
\makeatother

\usepackage[left=0.625in,right=0.625in,top=0.75in,bottom=1.05in]{geometry}
\begin{document}

\title{
%A Noise Resilient Approach using Active Learning for Keyword Spotting
Adaptive Noise Resilient Keyword Spotting Using One-Shot Learning}

\makeatletter
\newcommand{\linebreakand}{%
  \end{@IEEEauthorhalign}
  \hfill\mbox{}\par
  \mbox{}\hfill\begin{@IEEEauthorhalign}
}
\makeatother

\author{\IEEEauthorblockN{Luciano Sebastian Martinez-Rau, Quynh Nguyen Phuong Vu, Yuxuan Zhang, Bengt Oelmann and Sebastian Bader}
\IEEEauthorblockA{Department of Computer and Electrical Engineering \\ Mid Sweden University, Sundsvall, Sweden \\ luciano.martinezrau@miun.se} 
}

\maketitle
\copyrightnotice

\begin{abstract}
Keyword spotting (KWS) is a key component of smart devices, enabling efficient and intuitive audio interaction. 
However, standard KWS systems deployed on embedded devices often suffer performance degradation under real-world operating conditions.
Resilient KWS systems address this issue by enabling dynamic adaptation, with applications such as adding or replacing keywords, adjusting to specific users, and improving noise robustness.
However, deploying resilient, standalone KWS systems with low latency on resource-constrained devices remains challenging due to limited memory and computational resources.
This study proposes a low computational approach for continuous noise adaptation of pretrained neural networks used for KWS classification, requiring only 1-shot learning and one epoch.
The proposed method was assessed using two pretrained models and three real-world noise sources at signal-to-noise ratios (SNRs) ranging from 24 to -3~dB. 
The adapted models consistently outperformed the pretrained models across all scenarios, especially at SNR$\leq$18 dB, achieving accuracy improvements of 4.9\% to 46.0\%. 
%These results highlight the efficacy of the proposed methodology for deployment on resource-constrained devices.
These results highlight the efficacy of the proposed methodology while being lightweight enough for deployment on resource-constrained devices.
%Furthermore, we outline the development, challenges, and trade-offs involved in creating a resilient, standalone KWS system for resource-constrained devices, aiming to deliver a user-friendly experience.
\end{abstract}


% INTRODUCTION
\section{Introduction}
\label{s1}
Keyword spotting (KWS) enables voice-driven interactions with intelligent devices by detecting predefined words or phrases within an audio stream. 
Unlike large-scale speech recognition, KWS triggers specific actions, such as activating virtual assistants, controlling smartphones, wearables, and smart home devices, or initiating voice commands in automotive, healthcare, and security applications. 
This makes KWS a core component of modern consumer electronics.

Most state-of-the-art literature focuses on improving the accuracy of KWS systems, often relying on computationally intensive deep learning-based models that are not feasible for low-power devices~\cite{9665775,9785955,10631631}. 
However, deploying KWS on resource-constrained embedded systems, such as microcontrollers (MCUs) and low-power edge devices, presents significant challenges in terms of computational efficiency, memory constraints, and energy consumption~\cite{9707596}. 
Furthermore, standalone and autonomous KWS systems on resource-constrained devices must operate reliably without internet connectivity to protect user privacy, complicating their resilient design at both the firmware and hardware levels~\cite{ulkar2021ultra}. To address these challenges, the individual modules of a standalone KWS system should be carefully designed and optimized. 

%MODEL SYSTEM
\begin{figure}[t]
      \centering
      \includegraphics[width=1.0\columnwidth]{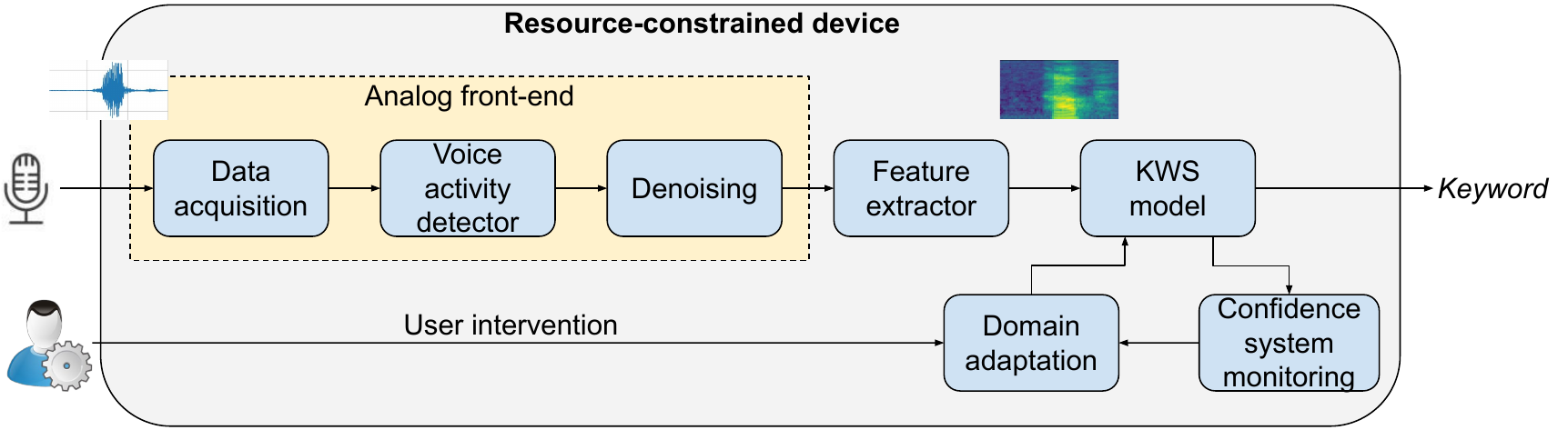}
      \caption{Block diagram of a standalone KWS.}
      \label{fig_block_diagram}
\end{figure}

The general structure of a typical KWS system is depicted in Fig.~\ref{fig_block_diagram}.
An analog front-end captures the acoustic signal, pre-processes it, and converts it to the digital domain.
The data acquisition module captures acoustic signals using a microphone, applying signal processing techniques to improve the signal-to-noise ratio (SNR) and automatic gain control for better distant speaker coverage~\cite{7178863}.
A voice activity detector minimizes unnecessary computation by enabling downstream processing only when speech is detected~\cite{7178970}.
Optional denoising techniques like beamforming or adaptive noise cancellation may further enhance speech clarity by mitigating reverberation, interference, and noise~\cite{7178863,10447074}.
The feature extractor transforms conditioned audio into discriminative representations, typically using time-frequency features such as log-Mel spectrograms or Mel-frequency cepstral coefficients~\cite{9707596,9233931}.
A machine learning KWS model, often based on efficient neural network (NN) architectures, classifies keywords. 
Model compression techniques such as quantization, pruning, and shrinking are commonly employed to reduce computation overhead~\cite{10136724}.
The confidence monitoring module assesses the reliability of KWS model predictions under uncertain conditions.
The domain adaptation module complements the previous module, enabling the system to adjust to changing operational conditions in real-time. It also allows external user intervention to tune the actual system configuration.

Recent advancements in adaptive KWS systems for embedded devices have addressed challenges such as adding new keywords defined by the user~\cite{michieli23_interspeech}, maintaining robustness against variations in speaker accents and pronunciations~\cite{cioflan2024boosting}, and mitigating background noise~\cite{10595987,Cioflan2022}.
These on-device noise adaptation methods work by contaminating clean samples stored on the device with captured noise signals to adapt a lightweight depthwise separable convolutional NN (DS-CNN).
While the performance of these methods improves as more samples are processed, this also leads to increased energy consumption and system latency, with a reported minimum latency of 14 seconds.

This study proposes a lightweight approach for continuous real-world noise adaptation in KWS using transfer learning (TL) on the last fully-connected layer of an NN model. The adaptation requires only one-shot learning and a single training epoch, significantly reducing data storage and computation requirements with respect to the state-of-the-art. This paper provides a proof-of-concept approach, which is designed to be deployable on resource-constrained devices in the future.

The remainder of the paper is organized as follows:
Section~\ref{s2} describes the state-of-the-art related to noise-resilient keyword spotting.
Section~\ref{s3} introduces the proposed method and dataset used in the experiments.
The results and their interpretations are given in Section~\ref{s4}. Finally, the study's conclusion and future direction are presented in Section~\ref{s5}.

\section{Related Works}
\label{s2}

Recent advancements in KWS have emphasized the development of lightweight deep learning models tailored for deployment on resource-constrained devices. DS-CNNs have emerged as a prominent architecture due to their balanced trade-off between accuracy, storage requirements, and computational efficiency. Notably, Zhang \textit{et al.} \cite{zhang2017hello} demonstrated the feasibility of deploying DS-CNNs on a Cortex-M7 STM32F746G MCU, achieving 95.4\% accuracy. 
Similarly, S{\o}rensen \textit{et al.}~\cite{sorensen2020depthwise} implemented a DS-CNN-based KWS system on an embedded platform, employing quantization techniques to reduce memory footprint without compromising performance.

While noise robustness has been extensively studied in automatic speech recognition~\cite{6732927}, its exploration within KWS, particularly on embedded systems, remains limited. 
Some studies have addressed this gap by enhancing noise robustness in deployed KWS solutions using on-site noise sources. 
For instance, Huang \textit{et al.}~\cite{8682682,8462346} applied speech enhancement techniques to improve "Ok Google" recognition in noisy environments. 
Jung \textit{et al.}~\cite{jung2020multi} introduced a multi-task network combining KWS and speaker verification, leveraging speech enhancement and feature extraction to bolster performance under noise. 
Park \textit{et al.}~\cite{park2021noisy} employed a knowledge distillation approach, enhancing noise robustness at the expense of increased inference latency and energy consumption. 
Wu \textit{et al.}~\cite{wu2020domain} proposed domain-aware training methods to improve far speaker distance KWS performance, addressing challenges posed by environmental noise and room reverberations.
The primary limitation of the aforementioned studies is that the deployed models cannot be adjusted to new operating conditions.

On-device learning has emerged as a promising approach for adapting KWS models to specific noise conditions encountered in real-world deployments.
Cioflan \textit{et al.}~\cite{Cioflan2022} proposed a domain adaptation methodology for KWS applications, achieving notable improvements in noise robustness. However, their approach incurred a high energy cost of 5.81\,J and lacked demonstration on MCUs, limiting its practicality for low-power devices. 
In a subsequent work~\cite{cioflan2024device}, the authors implemented a similar methodology on MCUs. Their approach involved storing noiseless utterances in the memory device and mixing them with real-time acquired noise to generate noisy utterances for model updates. While effective with substantial utterance data, the method's performance on MCUs is constrained by memory limitations. Moreover, experimental performance was evaluated under a specific noise condition (SNR=0~dB), lacking consideration of varying noise levels or SNR discrepancies between training and testing phases.

The current work addresses these limitations by presenting a proof of concept for adapting a pretrained NN to specific noise types using low memory and computational resources. This approach aims to enhance the practicality and deployability of KWS systems in real-world, resource-constrained environments by significantly reducing the amount of required on-site training data for the adaptive learning task.

\section{Noise Adaptation Approach}
\label{s3}
This section presents the proposed methodology for developing a voice command recognizer capable of operating in noisy environments.
%The approach focuses on designing NNs potentially able for resource-constrained devices and involves two key steps: training models for KWS classification, and using on-device learning to adapt these models to specific noise conditions once they are deployed. 
The approach performs continuous noise adaptation by fine-tuning pretrained NNs.
%It allows integration in an autonomous KWS system that integrates confidence monitoring.
Both clean and noisy audio signals are utilized during model development.

\subsection{Datasets}
The proposed approach employs the Google Speech Command (GSC) version~2 dataset, which contains 105,000~one-second speech utterances of 35~keyword classes and six speechless audio files with background noise~\cite{warden2018speechcommandsdatasetlimitedvocabulary}.
A 12-class benchmark is commonly used for KWS systems, including the following 10~keywords: "Yes", "No", "Up", "Down", "Left", "Right", "On", "Off", "Stop" and "Go". 
The remaining 25 keywords are grouped into an "Unknown" class, and a "Silence" class is created by randomly extracting one-second clips from the speechless recordings.

Clean one-second utterances from the GSC dataset are contaminated with additive noise from various sources to generate noisy signals.
The noise sources include: 
\begin{enumerate}
    \item Colored "White" and "Pink" noise.
    \item General indoor spaces noise categorized as "Babble"~\cite{VARGA1993247}, "Office", "Kitchen", and "Living room"~\cite{thiemann_2013_1227121}.
    \item Specific on-site noise produced by "Car horn", "Dog bark", and "Street music"~\cite{Salamon:UrbanSound:ACMMM:14}.
\end{enumerate}

%The clean and noisy audio signals, sampled at 16~kHz, are converted to the log-Mel time-frequency domain using 25~ms windows, a 10~ms hop size, and 64~Mel-filters spanning 50–7500~Hz. 
%This preprocessing step reduces the signal length by a factor of~2.48 and can be performed using an external low-power analog front-end, making it suitable for resource-constrained devices~\cite{9707596}.

\subsection{Method}

%MODEL FIGURE
\begin{figure}[t]
      \centering
      \includegraphics[width=1.0\columnwidth]{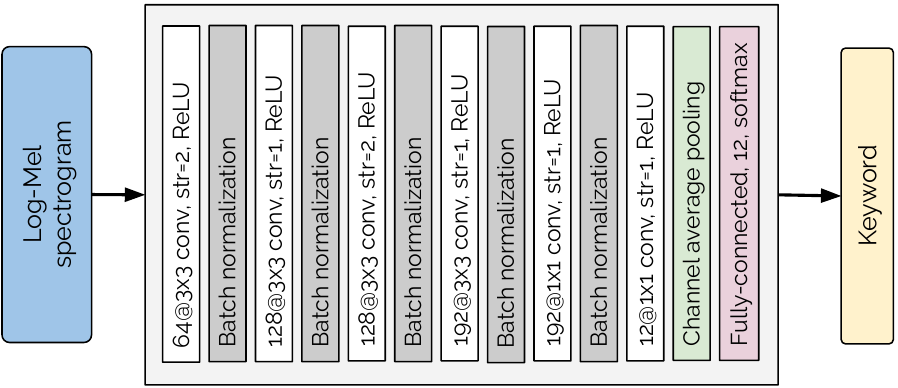}
      \caption{Proposed model architecture. Each convolution layer shows the number of kernels, kernel size, and activation function. The average pooling layer takes the average value from the whole image of each kernel. Fully-connected layer indicates the number of neurons and activation function.}
      \label{fig_model}
\end{figure}

The method described below first generates two pretrained models for KWS classification, which are then used to adapt to specific noise conditions.
The end-to-end classification of keywords using acoustic signals begins with the computation of the log-Mel spectrogram. 
In this study, we follow the same procedure as~\cite{9707596} to convert the acoustic signal to the time-frequency domain. 
The clean and noisy audio signals, sampled at 16~kHz, are converted to the log-Mel time-frequency representation using 25~ms windows, a 10~ms hop size, and 64~Mel-filters spanning 50–7500~Hz. 
This preprocessing step reduces the signal length by a factor of~2.48.
As demonstrated in~\cite{9707596}, this log-Mel spectrogram configuration can be computed using an external low-power analog front-end, significantly reducing the energy consumption of an end-to-end keyword spotting system on resource-constrained devices.

The proposed model architecture, presented in detail in Fig.~\ref{fig_model}, takes the log-Mel spectrogram as input to classify one of the twelve keyword classes.
Inspired by the CNN model used in~\cite{9707596}, our architecture comprises five convolutional layers interspersed with batch normalization layers, followed by an average pooling layer that computes the mean across the entire image channel. 
This single-channel value is then processed by a fully-connected layer.

The architecture is used to train two models. 
The first, named \emph{baseline-model}, is trained exclusively on clean voice command data and serves as a reference for optimal noise conditions. 
The second, called \emph{noise-aware-model}, is trained using both clean and noisy data, attempting to enhance robustness in adverse noise conditions.

To enable adaptation to continuous noise conditions, this study explores the use of few-shot TL techniques. 
Only the last fully-connected layer is fine-tuned, keeping all convolutional layers frozen to minimize memory and computational requirements.
This technique allows for adjusting the pretrained models to changing noise scenarios over time, improving their robustness in shifting environments.
In the future, the pretrained model and the continuous noise adaptation are intended to be deployed on resource-constrained devices for real-time operation. Therefore, a lightweight implementation is key.

\subsection{Experimental Setup}
The GCS dataset with clean utterances is divided into train, validation, and test sets in an 80:10:10 ratio.
Each set contains audio clips balanced by class, as specified in the original dataset. 
Additionally, the audio clips are mixed with individual noise sources to create noisy train, validation, and test sets with SNRs ranging from \mbox{-3} to 24~dB in 3~dB increments.
Model performance is quantified using the accuracy metric.

The first experiment compares the two pretrained models.
The \emph{baseline-model} is trained on the clean training set and evaluated on both clean and noisy test sets, with the noise including colored and indoor noise sources.
Five instances of the \emph{noise-aware-model} are trained using the clean training set combined with an additional~20, 40, 60, 80, or~100\% of a balanced noisy training set.
The noisy training dataset includes an equal number of samples contaminated by each noise source and specific SNR level.
These instances are evaluated similarly to the \emph{baseline-model}.

In this experiment, the Adam optimizer and cross-entropy loss function are used for training, with an initial learning rate of~$10^{-4}$.
The learning rate decays by a factor of~$10^{-1}$ if no improvement is observed for 10 consecutive epochs.
Training is conducted with a batch size of 16 and concludes after 50~epochs. Default values are retained for all other parameters.
The experiment is repeated 25 times using different model weight initializations to obtain reliable results.

The second experiment evaluates the domain adaptation of the pretrained \emph{baseline-model} and the best-performing \emph{noise-aware-model} to specific on-site noise sources by applying few-shot TL.
%To minimize memory and computational requirements, as discussed in Section~\ref{s2}, only the fully-connected layer is fine-tuned using transfer learning techniques, keeping all other layers frozen. 
This experiment investigates the number of training samples and epochs required for adaptation, as well as the models' robustness to changing SNR conditions after adaptation.
The number of samples per class (shots) is evaluated between~1 and~5, varying the number of training epochs from~1 to~5 for each case.
During adaptation, backpropagation computes the stochastic gradient descent with a learning rate of~$10^{-4}$ and cross-entropy loss function.

All code was developed using Python 3.12.8\footnote{\scriptsize{Code available at: https://github.com/lucianomrau/NoiseRobust\_KeywordSpottingMCU}}. 
Audio signal processing was performed using the Torchaudio~2.5.1 library, while model architecture definition, training, and testing were implemented using Pytorch~2.5.1.
Experiments were conducted in an Intel Core™ i9-12900K CPU with 32~GB RAM and a NVIDIA GeForce RTX~3080 GPU with 10~GB memory.

%FIGURE HEATMAP
\begin{figure}[t]
      \centering
      \includegraphics[width=1.0\columnwidth]{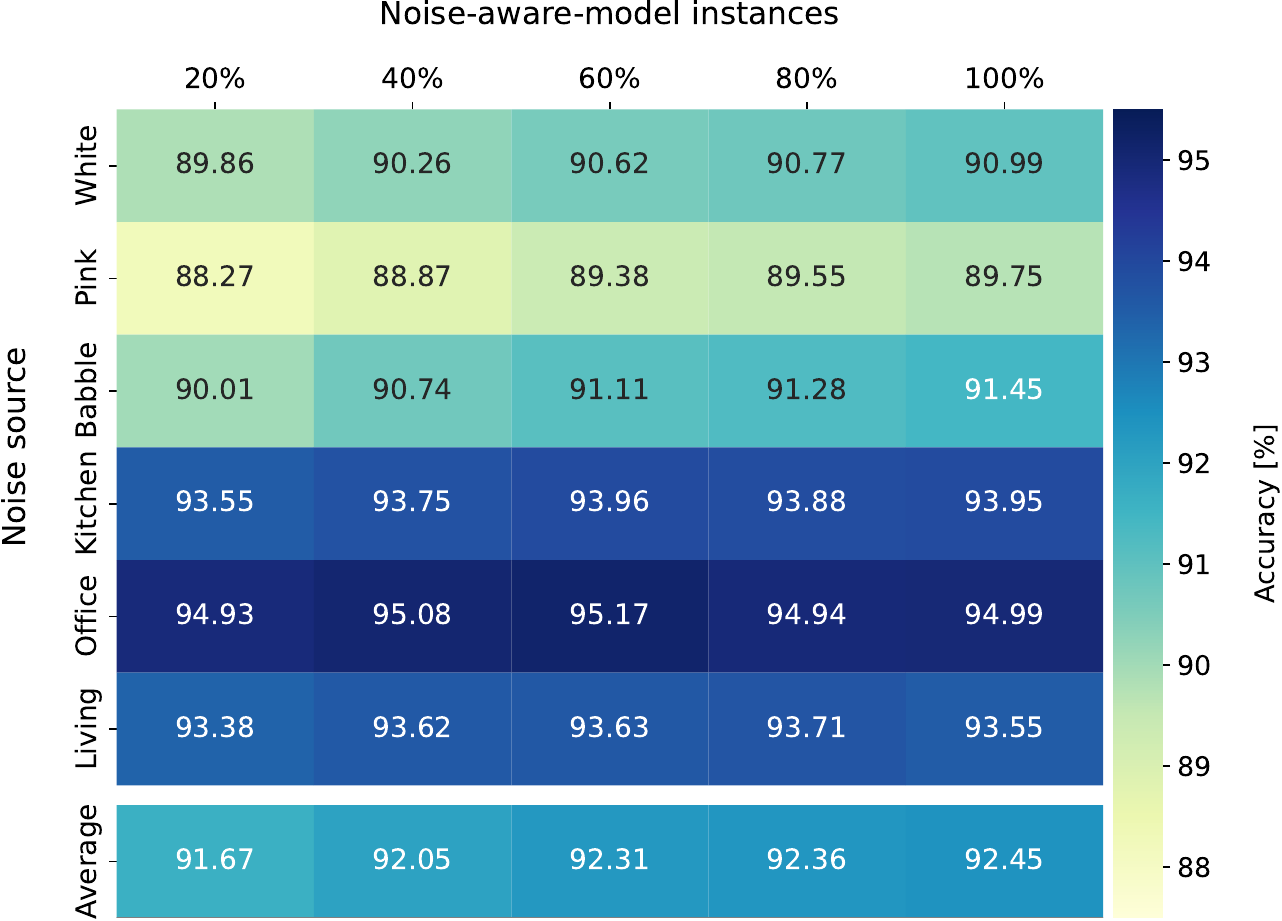}
      \caption{Performance of \emph{noise-aware-model} instances evaluated across different noise sources. Performance is measured as the mean classification accuracy across all trained models (25), evaluated at different SNR levels for the corresponding noise source.}
      \label{fig_noise_aware_performance}
\end{figure}

\section{Results and Discussion}
\label{s4}

\subsection{Comparison of pretrained models}
In the initial experiment, both the \emph{baseline-model} and each of the five instances of the \emph{noise-aware-model} are trained 25 times with different weight initializations. The average accuracy over these 25 repetitions is used to evaluate the model's performance.
The \emph{baseline-model}, trained with noiseless utterance samples, achieves a mean accuracy of 96.27\% for classifying clean (noise-free) keyword utterances. 
% In comparison, the \emph{noise-aware-model} instances, trained with clean signals plus 20\%, 40\%, or 60\% additional noisy signals, achieve similar accuracies of 96.26\%, 96.42\%, and 96.11\%, respectively. This represents a decrease of less than 0.22\% compared to the \emph{baseline-model}. 
% However, the accuracy on a clean test set decreases by 1.06\% and 0.49\% for the \emph{noise-aware-model} instances trained with an additional 80\% and 100\% of noisy signals, respectively.
In comparison, the \emph{noise-aware-model} instances, trained with clean signals plus 20\%, 40\%, 60\%, 80\%, and 100\% additional noisy signals, achieve a declining accuracies of 96.04\%, 95.98\%, 95.95\%, 95.81\%, and 95.73\%, respectively. This represents a decrease of less than 0.54\% compared to the \emph{baseline-model} (p\lt0.001 in all comparisons; Wilcoxon signed-rank test~\cite{Wilcoxon1945-zq}). 

Figure~\ref{fig_noise_aware_performance} presents the mean accuracy of the \emph{noise-aware-model} instances when tested with different noise sources (colored and indoor noise) at SNRs ranging from \mbox{-3} to 24~dB in 3~dB increments.
% While no single optimal level of added noise during training achieves the best performance across all noise sources, the \emph{noise-aware-model} instance trained with 100\% noise attains top accuracy for 3 of 6 noise sources and the highest overall mean accuracy. 
% "Statistical validation using Friedman tests revealed significant performance differences between models ($\chi^{2}=1126.856$). 
% Post-hoc Wilcoxon signed-rank tests with Holm correction confirmed that the \emph{noise-aware-model} instance trained with 100\% noise significantly outperformed all others (all adjusted p\lt0.001).
% Based on these results, the 100\% noise instance is adopted in the remainder of this work for continuous noise adaptation.
While no single noise level added during training yields universally optimal performance across all noise sources, the 100\% noise-trained \emph{noise-aware-model} demonstrates superior accuracy for 3 of 6 noise sources and the highest overall mean accuracy. 
Statistical validation confirms this observation, with Friedman tests indicating significant performance variation between models ($\chi^{2}$=1126.856, p\lt0.001) and post-hoc Wilcoxon signed-rank tests (Holm-corrected) showing the 100\% noise variant significantly outperforming all alternatives (all p\lt0.001). Consequently, the \emph{noise-aware-model} instance trained with 100\% noise is selected as the optimal \emph{noise-aware-model} for subsequent continuous noise adaptation experiments.

Figure~\ref{fig_general_noise_performance} compares the noise robustness of the \emph{baseline-model} and the selected \emph{noise-aware-model} for specific colored and indoor noise sources at varying SNRs.
As expected, lower SNRs lead to reduced performance for both models, regardless of the noise source. 
Notably, the \emph{noise-aware-model} consistently outperforms the \emph{baseline-model} across all cases, with the most significant improvements observed for colored noise sources.
%This demonstrates that improved generalization of the \emph{noise-aware-model} is due to two factors: a larger training dataset and the presence of similar noise patterns in both the training and test sets.
This demonstrates that the \emph{noise-aware-model} learns from the noise patterns added to the training set.

%FIGURE COMPARATIVE PERFORMANCE
\begin{figure}[t]
      \centering
      \includegraphics[width=0.8\columnwidth]{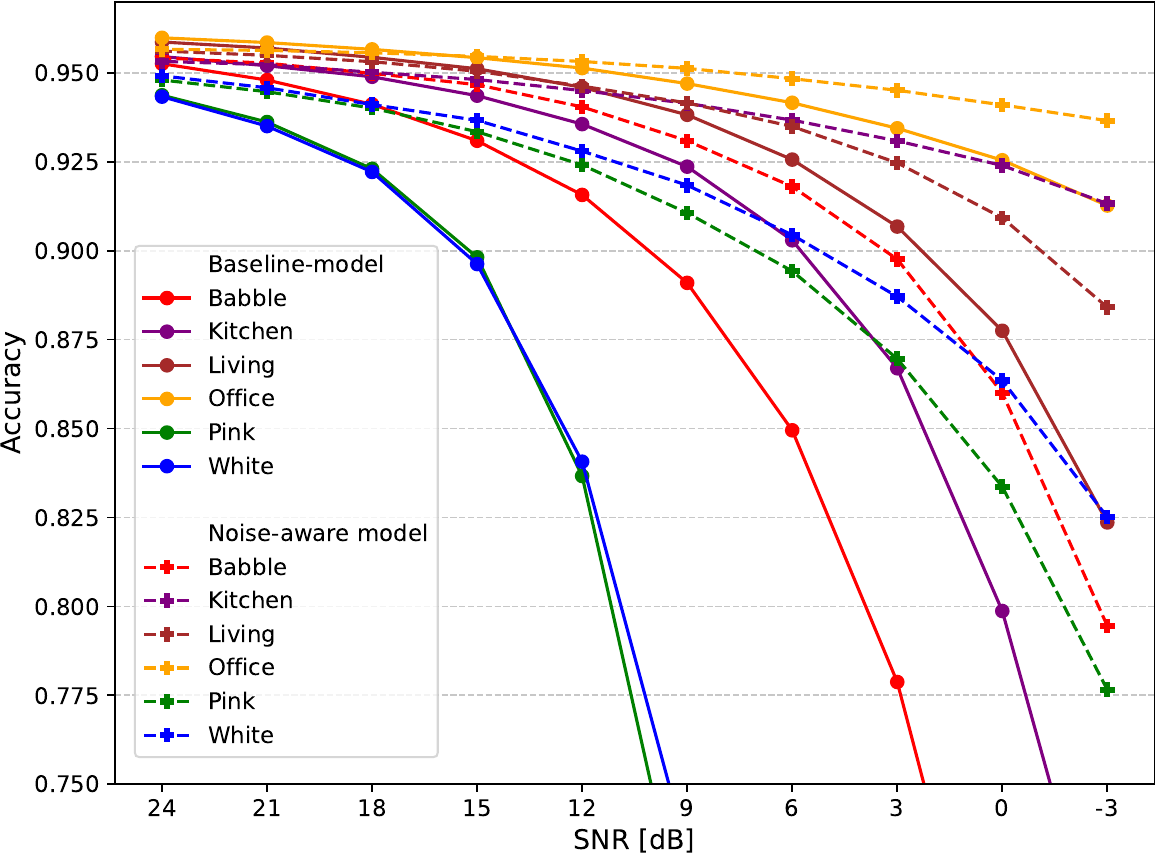}
      \caption{Comparative performance for the \emph{baseline-model} and the selected \emph{noise-aware-model} when tested across different noise sources and SNRs, averaged across all trained models.}
      \label{fig_general_noise_performance}
\end{figure}

%FIGURE 1-SHOT
\begin{figure*}[th!]
      \centering
      \includegraphics[width=0.95\textwidth]{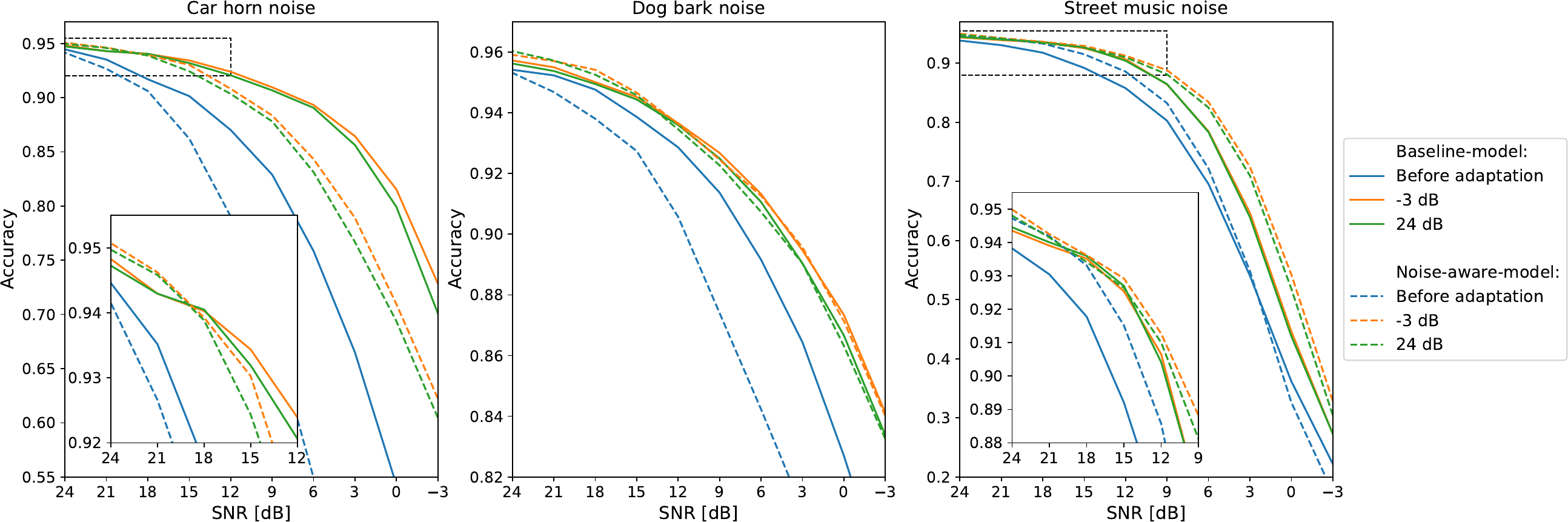}
      \caption{Tested models in specific noise sources before and after training with 1-shot for 1 epoch.}
      \label{fig_one_shot}
\end{figure*}

%FIGURE EPOCHS
\begin{figure*}[th!]
      \centering
      \includegraphics[width=.9\textwidth]{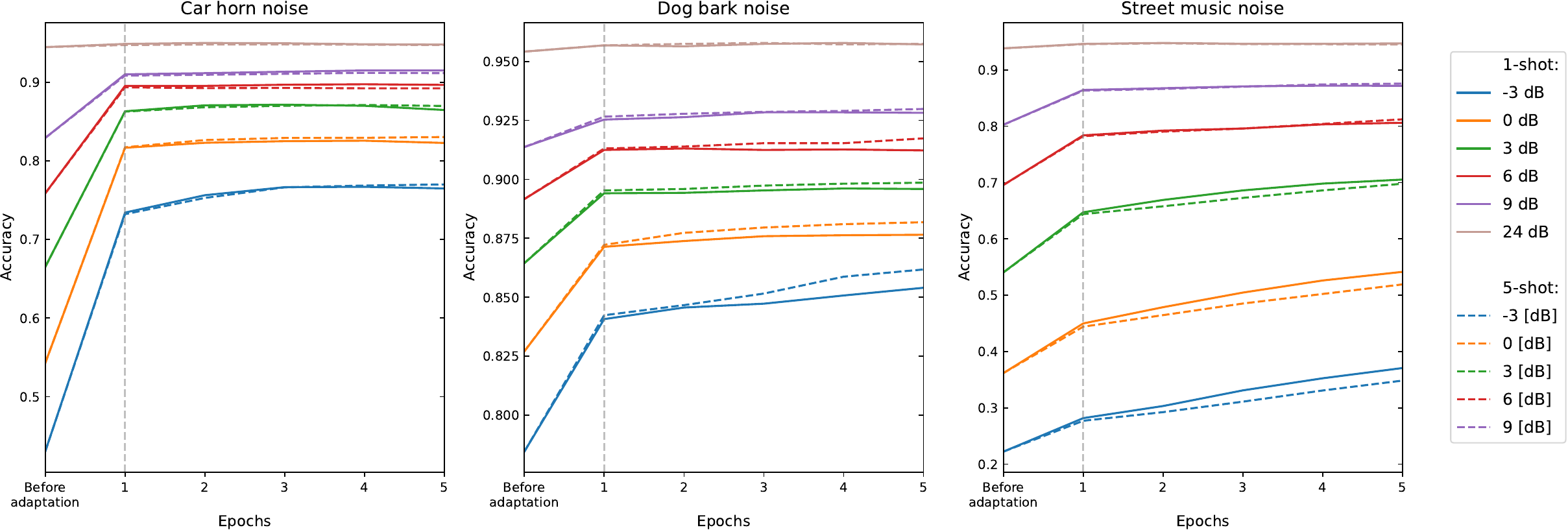}
      \caption{Accuracy of the pretrained \emph{baseline-model} before and after continuous adaptation for 1 to 5 epochs for different on-site noise conditions. Adaptation is performed using 1- or 5-shot per epoch.}
      \label{fig_epochs}
\end{figure*}

\subsection{On-site noise adaption}
Both pretrained models are adapted to a specific on-site noise condition (source and SNR value) using 1-shot learning and 1~epoch to ensure minimal storage and processing requirements. 
Figure~\ref{fig_one_shot} shows the results of adapting the pretrained models at minimal and maximal SNRs (-3 and 24~dB, represented by orange and green lines, respectively). 
The adapted models are tested across the entire SNR range to account for the non-stationary nature of dynamic real-world noise conditions.
For models adapted to intermediate SNR values (-3~dB$\lt$SNR$\lt$24~dB) not shown in Fig.~\ref{fig_one_shot}, the accuracy curves consistently fall within the range delineated by the \mbox{-3} and 24~dB curves.
The results indicate that the pretrained models effectively adapt to new conditions, achieving superior performance and robustness across different SNRs when compared to the original models (blue lines in Fig.~\ref{fig_one_shot}). 
The best-performing adapted model depends on the specific noise scenario.
Models adapted at \mbox{-3~dB} achieve higher accuracy at low and middle range (SNR$\lesssim$18~dB) than those adapted at 24~dB. 
In contrast, at SNR$\gtrsim$18~dB, the adapted \emph{noise-aware-model} attains the highest accuracy, regardless of whether it was trained at \mbox{-3} or 24~dB.
However, at higher SNRs, the pretrained models before adaptation already exhibit accuracies above~90\% for all noise sources, raising questions regarding the necessity of adaptation.
Therefore, the conditions for adapting a continual learning model should be carefully selected.

When comparing the original pretrained models before adaptation (solid and dashed blue lines in Fig.~\ref{fig_one_shot}), the \emph{baseline-model} outperforms the \emph{noise-aware-model}, exhibiting higher noise robustness in two of the three noise sources. 
The robustness gap between these models is smaller for the "Street music" noise source.
However, the performance of both models is significantly affected by this noise source. 
These results reveal that the \emph{noise-aware-model} did not improve its generalization ability when training data containing colored and general indoor noise was added. 
Therefore, it does not offer a clear advantage over the \emph{baseline-model} for dynamic noise adaptation.

For large memory and computational devices, increasing the number of training samples and epochs may yield improvements in adaptation.  
Figure~\ref{fig_epochs} shows the performance variations as a function of the increasing number of epochs used for adapting the pretrained \emph{baseline-model}. 
The model is trained for~1 to 5~epochs with 1- or 5-shot learning utilizing utterances contaminated with specific noise sources and SNRs.
Model evaluation is done at the same noise source and SNR value used during training adaptation.
The results show that the model adapts well to noise, increasing its performance when trained for only 1 epoch.
Adaptation capabilities can still be improved by increasing the number of training epochs, particularly for highly contaminated signals with noise (SNR$\lt$3~dB), and in some cases for values between 3 and 9~dB. 
The performance remains relatively constant for SNRs higher than 9~dB. Note that for clarity reasons, we have chosen not to show all SNR conditions in Fig.~\ref{fig_epochs}.

The effect of increasing the number of training samples from 1-shot to 5-shot learning presents similar accuracy when adaptations are done for 1 epoch.
Performance differences between 1-shot and 5-shot continual learning are observed as the number of epochs increases, especially for SNR$\lt$3~dB.
For the "Car horn" noise source, there is no clear improvement when adding more data during adaptation, while for "Street music", 1-shot learning achieves higher accuracy. 
For the "Dog bark" noise source at SNR$\leq$6~dB, 5-shot learning performs better than 1-shot, particularly as the number of training epochs increases.
However, it is important to note that 5-shot learning requires storing and processing five times more data per epoch compared to 1-shot learning.
Similar patterns are observed in the \emph{noise-aware-model} when evaluating performance by increasing the number of epochs and data used during adaptation.

%% CONCLUSION
\section{Conclusion}
\label{s5}
%This study addressed the challenges and trade-offs of deploying resilient KWS systems on resource-constrained devices, taking performance, computational efficiency, and user experience into account. 
%A key requirement for resilient KWS systems is their ability to dynamically adapt to real-world noise conditions.
A key requirement of real-time KWS systems operation is their ability to adapt to changing noise conditions.
To this end, this study proposed a few-shot TL approach for continuous noise adaptation of pretrained NNs.
We developed two pretrained models in a first experiment, a \emph{baseline-model} trained on noiseless utterances and a \emph{noise-aware-model} trained on a mix of noiseless and noisy utterances contaminated with colored and general indoor noises.
The continuous noise adaptation capabilities of both pretrained models were rigorously evaluated under three specific on-site noise sources at different SNRs. 

Experimental results demonstrated that both pretrained models significantly improved their performance after adaptation, with greater performance improvements observed as more noise is present.
While the \emph{noise-aware-model} outperforms the \emph{baseline-model} when tested on colored and general indoor noises before adaptation, the best-performing pretrained model after noise adaptation depends on the noise source.

The effect of varying the number of shots and epochs during adaptive training was also analyzed, demonstrating that good performance can be achieved with only 1-shot and 1 epoch.
Increasing the number of epochs from 1 to 5 leads to higher accuracy when the signals are very noisy (low SNR values), but comes at a cost of increasing the computation and therefore the system latency.
Increasing the number of shots per epoch from 1 to 5 also does not guarantee performance improvement in all cases, but requires storing and processing more data.

Given the proposed approach's low computational requirement, future work will focus on deploying it on resource-constrained devices.
Moreover, future work will focus on automatically detecting when to perform the adaptation, which is a key requirement for effective on-device KWS adaptation.

\section*{Acknowledgement}
The authors would like to acknowledge financial support of this research by the Knowledge Foundation under grant number 20180170 (NIIT).

\balance
\bibliographystyle{./IEEEtran}
%\bibliography{references_abbrev}
\bibliography{references}

\end{document}